\documentclass{appolb}

\usepackage{graphics}
\usepackage{epsfig}
\usepackage[latin1]{inputenc}
\usepackage{cite}

\newcommand{\dg}{\ensuremath{^{\circ}}}
\def\xbj{\ensuremath{x_{Bj}}}
\newcommand{\gev}{\,\mbox{GeV}}

\newcommand{\gevsq}{\ensuremath{\mathrm{GeV}^2} }

\def\prp{\perp}
\def\prp{t}

\def\sx{small-$x$}
\def\kt{\ensuremath{k_\prp}}
\def\kti#1{\ensuremath{k_{\prp #1}}}

\def\pt{\ensuremath{p_\prp}}
\def\pti#1{\ensuremath{p_{\prp #1}}}

\def\qbi#1{\ensuremath{\bar{q}_{ #1}}}

\def\cascade{{\sc Cascade}}

\def\ldcmc{{\sc Ldc}}

\def\as{\ensuremath{\alpha_s}}
\def\sub#1{\ensuremath{_{\mbox{\scriptsize #1}}}}
\def\alb{\ensuremath{\bar{\alpha}\sub{s}}}

\newcommand{\CCFM}{CCFMa,CCFMb,CCFMc,CCFMd}
\newcommand{\BFKL}{BFKLa,BFKLb,BFKLc}
\newcommand{\LDC}{LDCa,LDCb}
\newcommand{\LDCMC}{LDCc,LDCd}
\newcommand{\alphasb}{\alb}

\newcommand{\DGLAP}{DGLAPa,DGLAPb,DGLAPc,DGLAPd}

\newcommand{\CASCADEMC}{jung_salam_2000,CASCADEMC}

\begin{document}
\title{Recent results from CCFM evolution%
}
\author{H. Jung
\address{  
Experimental High Energy Physics\\
 Department of Physics, Lund~University, 
SE-22100~Lund,~Sweden }
}

\maketitle
\begin{abstract}
Recent developments of the small $x$  CCFM evolution 
are described, including improvements of the splitting function. The
resulting unintegrated gluon densities are used for predictions of hadronic
final state measurements like jet production at HERA and heavy quark production
at the Tevatron.
\par
This paper is dedicated to the memory of Jan Kwiecinski, 
who passed away so early. 
\end{abstract}

\section{The problem with collinear factorization}
Inclusive cross section measurements~\cite{H1F2,ZEUSF2}, but also many hadronic
final state cross sections
\cite{H1alpha,ZEUSalpha,D0comp}
are in general well described in QCD. 
The cross sections are factorized into integrated parton density functions 
(PDFs) which are evolved
from a starting scale $\mu_0$ to the factorization
 scale $\mu_f$ using  the DGLAP~\cite{\DGLAP} evolution
equations and  convoluted with the process dependent coefficient functions (matrix
elements). This approach is often called collinear factorization.
\par
However, regions in phase space 
 exist where the above mentioned approach using coefficient
functions evaluated at fixed order in the strong coupling constant, $\as$, is
not sufficient to describe the measured cross sections. 
Such measurements are e.g.
forward jet production and  the azimuthal de-correlation of jets at HERA, but
also heavy quark production at the Tevatron.
\par
Processes sensitive to hard multi-parton emissions are generally 
not covered fully in DGLAP. The BFKL~\cite{\BFKL} approach, 
also called the \kt-factorization~\cite{CCH,CE} or the semi-hard
approach~\cite{GLR,LRSS2}, is assumed to be better suited for such a scenario,
since partons in the initial state cascade can have any kinematically allowed
transverse momentum, in contrast to DGLAP.
The CCFM~\cite{\CCFM} approach and its reformulation in the Linked Dipole Chain
Model LDC~\cite{\LDC}
attempt  to cover both the DGLAP and BFKL
regions by considering color coherence effects.
A general introduction to \sx\ physics and the evolution equations 
can be found  in~\cite{smallx_2001}.

\section{CCFM - LDC evolution and unintegrated parton densities}
In the CCFM evolution equation  angular
ordering of emissions is introduced
to correctly treat gluon coherence effects. In
the limit of asymptotic energies, it is almost equivalent to 
BFKL~\cite{Forshaw:1998uq,Webber:1998we,Salam:1999ft}, but 
also similar to the DGLAP evolution for large $x$ and high
$Q^2$.  The cross section is \kt-factorized into an off-shell
matrix element convoluted with an unintegrated parton density (uPDF), which
now also contains a dependence on the maximum angle $\Xi$ allowed in
emissions.
This  maximum allowed angle $\Xi$
is defined by the hard scattering quark box, producing the 
(heavy) quark pair and also defines the scale for which parton emissions
are factorized into the uPDF.
\par
The original CCFM splitting function is given by:
\begin{equation}
P_g= \frac{\alphasb(\pt)}{1-z} + 
\frac{\alphasb(\kt)}{z} \Delta_{ns}\; \;\;\mbox{with }\; \log\Delta_{ns} =
  -\int_{z_i}^1 \frac{dz'}{z'} 
\int^{\kti{i}^2}_{(z'\qbi{i})^2} \frac{d q^2}{q^2}\alphasb		  
\label{Pgg}
\end{equation}
with $\alphasb=\frac{3 \as}{\pi}$, 
The phase space region of angular ordered emissions is obtained from:
\begin{equation}
z_{i-1} \qbi{i-1} < \qbi{i} \; \;\;\mbox{with }\; 
\qbi{i}  = \frac{\pti{i}}{1-z_i}
\label{ang_ord}
\end{equation}
Here $z_i = x_i /x_{i-1}$ is the ratio of the energy fractions 
in the branching $(i-1) \to i$ and $\pti{i}$ is
the transverse momentum of the emitted gluon $i$. The transverse momentum
of the propagating gluon is given by $\kti{i}$.
It is interesting to note, that the angular ordering constraint, as given by
eq.(\ref{ang_ord}), reduces to ordering in transverse momenta 
$\pt$ for large $z$,
whereas for $z\to 0$, the transverse momenta are free to perform a 
so-called random walk.  
\par
The  LDC model is a
reformulation of CCFM, where the separation between the initial- and
final-state emissions is redefined and, in addition to the angular
ordering in eq.(\ref{ang_ord}), the gluons emitted in the initial state are
required to have
\begin{equation}
  p_{\perp i} > \min(k_{\perp i},k_{\perp i-1}).
    \label{ldccut}
\end{equation}
This constraint  means that the $\pt$ of the
emitted gluon is always close to the highest scale in the splitting
and the argument in $\as$ is naturally taken to be
$\pt$. 
\par
While formally equivalent to the DLLA accuracy, 
it is important to note that the  
sets of chains of initial-state splittings summed over, are
different in LDC and CCFM. One chain in LDC corresponds to a whole
set of chains in CCFM, and predictions for non-inclusive observables
of the evolution will only be comparable if the correct final-state
emissions are added in the regions relevant for the formalism used.
\par
The \cascade~\cite{\CASCADEMC} program is a direct implementation of the  
CCFM approach into a hadron level Monte Carlo generator using a backward
evolution approach for the initial state parton shower. 
The \ldcmc~\cite{\LDCMC} program is a
realization of the LDC model in terms of a hadron level Monte Carlo generator.
\subsection{Unintegrated gluon densities}\label{ccfmfits}
The original CCFM splitting function given in eq.(\ref{Pgg}) includes only the
singular terms as well as a simplified treatment of the scale in $\as$.
Due to the angular ordering 
a kind of random walk in the propagator gluon 
\kt\ can be performed, and therefore care has to be taken 
for small values of \kt. 
For values of $\kt < \kt^{cut}$ the non-perturbative region is entered, which is
avoided in a strictly \pt-ordered evolution (DGLAP). In CCFM, for 
$\kt < \kt^{cut}$, $\as(\kt)$ and also the 
splitting probability could become un-physically large.  
However, this is the region of high parton densities, where saturation
effects could appear. A practical treatment here is to freeze $\as$ and the
gluon density for values  $\kt < \kt^{cut}$.
\par
Three new sets of
uPDFs were determined~\cite{jung-dis03}:
{\it J2003 set 1} with the splitting function given in eq.(\ref{Pgg}), 
{\it J2003 set 2} including also the non-singular terms in $P_g$ and 
{\it J2003 set 3} using \pt\ as the argument in \as.
For all sets the input parameters were fitted to describe  
the structure function $F_2$ as measured at 
H1~\cite{H1_F2_1996,H1_F2_2001} and ZEUS~\cite{ZEUS_F2_1996,ZEUS_F2_2001}
in the range of $x < 10^{-2}$ and  $Q^2 > 5$~GeV$^2$. Using 248 data points
a $\chi^2/ndf = 1.29, 1.18, 1.83$ for {\it J2003 set 1,2,3 },
respectively, is obtained. 
\begin{figure}[tb]
\begin{center}
\resizebox{1\textwidth}{!}{\includegraphics{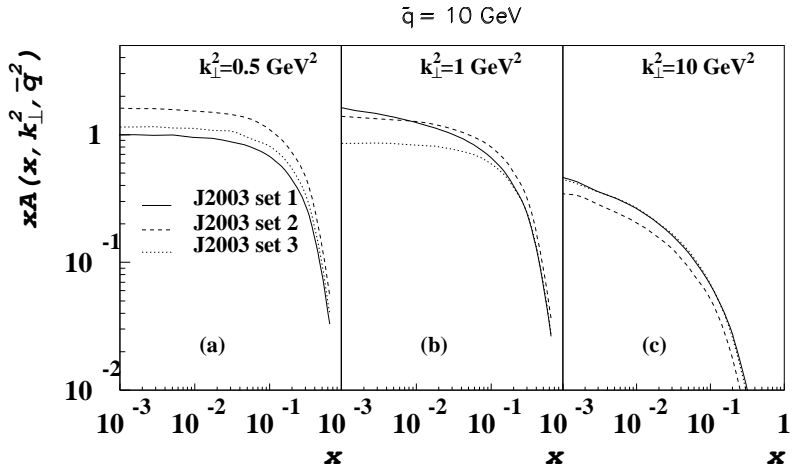}} 
 \caption{
 {\it Comparison of the different sets of unintegrated gluon densities obtained
 from the CCFM evolution  as a function of $x$ for different values of \kt\ at a
 scale of $\bar{q}=10$ GeV. 
 \label{ccfm-new}}}
\end{center}
\end{figure}
A comparison of the different sets of CCFM uPDFs is shown
in Fig.~\ref{ccfm-new}. 
\begin{figure}[tb]
\begin{center}
\resizebox{0.9\textwidth}{!}{\includegraphics{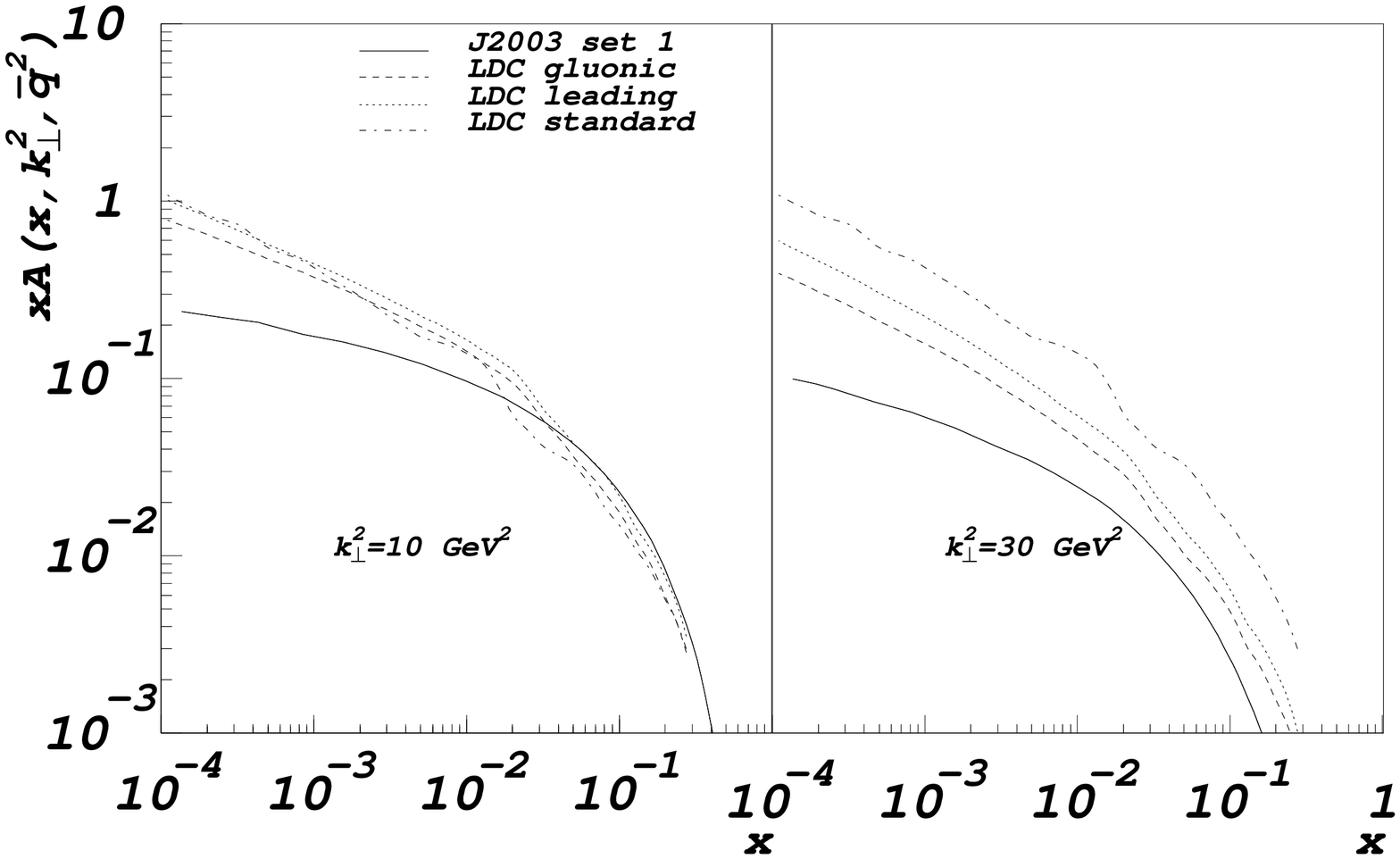}} 
 \caption{
 {\it Comparison of the different sets of unintegrated gluon densities obtained
 within LDC at
 scale of $\bar{q}=10$ GeV. \textrm{standard} refers to the full
 LDC including quarks in the evolution and the full gluon splitting
 function. For \textrm{gluonic} and \textrm{leading} only gluon
 evolution is considered with only singular terms in the splitting 
 function for the latter. Also shown is the { J2003 set 1} 
 for comparison (divided by $\pi$).
 \label{ccfm-ldc}}}
\end{center}
\end{figure}
\par
Also the LDC model describes $F_2$ satisfactorily well, but the
corresponding unintegrated gluon densities are somewhat different,
as also quarks can be included in the
evolution. In Fig.~\ref{ccfm-ldc} three different unintegrated
gluon densities for the LDC approach are presented.
The \textit{standard} set refers to the full
LDC including quarks in the evolution and the full gluon splitting
function, whereas for the \textit{gluonic} set and the 
\textit{leading} set only gluon
evolution is considered with only singular terms in the splitting 
function for the latter. All three alternatives have been
individually fitted to $F_2$ in the region $x<0.3$,
$Q^2>1.5$~GeV$^2$ for \textit{standard} and $x<0.013$ and
$Q^2>3.5$~GeV$^2$ for \textit{gluonic} and \textit{leading}.   
In LDC there is only one relevant infrared cutoff, $\kti{0}$,
which limits the $\pt$ of emitted gluons. 
\section{Comparison with hadronic final state data}
A comparison of measurements of hadronic final
state properties, like jet or heavy quark cross sections, with theoretical
predictions requires Monte Carlo event generators,
which also allow to apply the hadronization step. 
In the following sections a few examples are presented where data are compared
to predictions obtained with \cascade\ based on the CCFM unintegrated gluon
densities, described in section ~\ref{ccfmfits}.
\subsection{Jet cross section at HERA}
\begin{figure}[ht]
\begin{minipage}{0.48\textwidth}
\vskip -1.0cm
\centerline{\resizebox{1\textwidth}{!}{\includegraphics{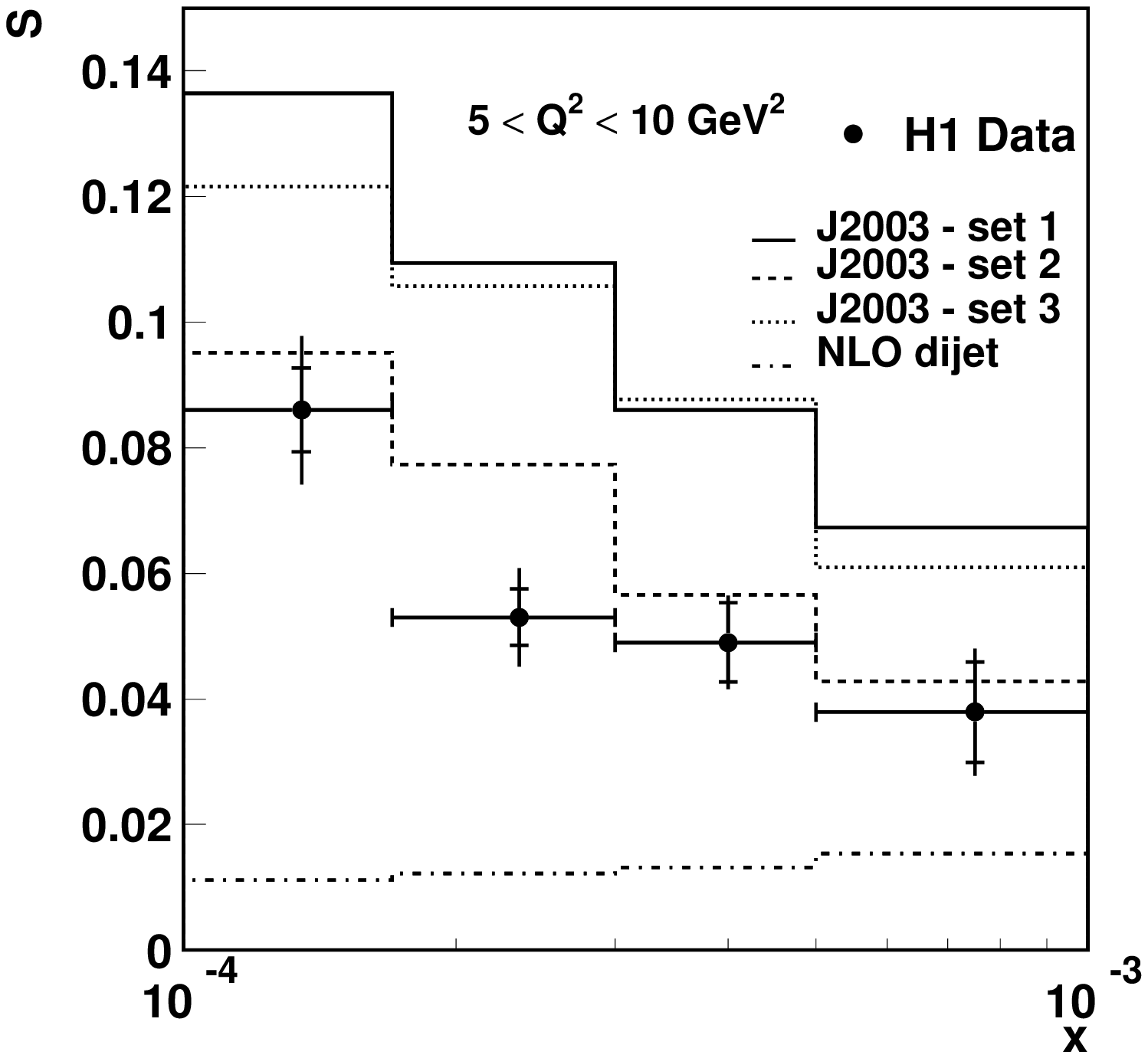}}} 
\vskip -0.2cm
\caption{{ The $S$ distribution for dijet events\protect\cite{H1dijet}. 
 The prediction from 
\cascade~  are shown together
with the NLO-dijet prediction in the collinear factorization approach.
   }\label{azim:casc-nlo}}
\end{minipage}
\hspace*{0.3cm}
\begin{minipage}{0.48\textwidth}
\vskip -1.0cm
\centerline{\resizebox{1\textwidth}{!}{\includegraphics{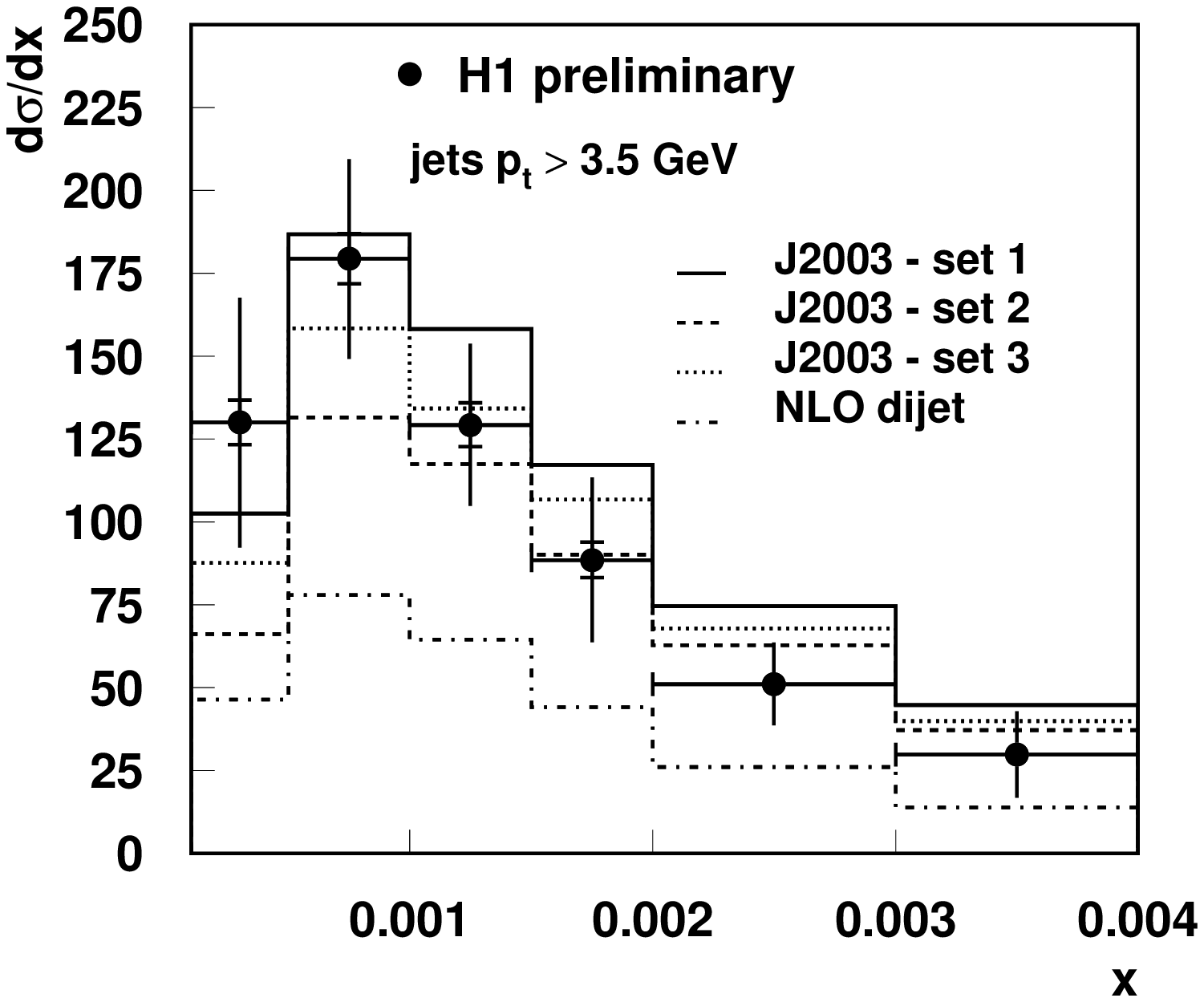}}} 
\vskip -0.5cm
\caption{The cross section for forward jet production as measured by 
H1\protect\cite{H1fwdjet2} as a function of \xbj. The prediction are the same as
in Fig.~\protect\ref{azim:casc-nlo}.
\label{fjet:cas-nlo} } 
\end{minipage}   
\end{figure}

The azimuthal correlation of dijets at HERA is sensitive
to the transverse momentum of the partons incoming to the hard scattering
process and therefore sensitive to the details of the unintegrated gluon 
density. This was studied in a measurement~\cite{H1dijet} of 
the cross section for dijet production with $E_T > 5 (7)$ GeV in the
range $1 < \eta_{lab} < 0.5$ in deep-inelastic scattering 
 ($10^{-4} < x < 10^{-2}$, $ 5 < Q^2 < 100$ GeV$^2$). 
In LO collinear factorization,  dijets at small
$\xbj$ are produced essentially by 
$\gamma g \to q\bar{q}$, with the gluon collinear to the
incoming proton. Therefore the $q\bar{q}$ pair is produced back-to-back in the
plane transverse to the $\gamma^* p$ direction. 
From NLO (${\cal O}(\alpha_s^2)$) on,  significant deviations
from the back-to-back scenario can be expected. In the \kt-factorization 
approach the transverse momentum
of the incoming gluon, described by the unintegrated gluon density,
is taken explicitely into account, resulting in deviations from a
pure  back-to-back configuration. 
The azimuthal de-correlation, as suggested in Ref.~\cite{Szczurek}, can be
measured:
\begin{equation}
 S = \frac{\int_{0}^{\alpha}N_{2-jet}(\Delta\phi^{*},x,Q^2)d\Delta\phi^{*}}
            {\int_{0}^{180^{\dg}}N_{2-jet}(\Delta\phi^{*},x,Q^2)d\Delta\phi^{*}}, 
                0 < \alpha < 180^{\dg} 
\end{equation}
In the measurement shown in Fig.~\ref{azim:casc-nlo},
$\alpha = 120^{\dg}$ has been chosen. The data are compared to 
predictions from \cascade\ using {\it J2003 set 1 - 3}. 
Also shown for comparison is the NLO-dijet~\cite{DISENT} calculation of the
collinear approach. 
One clearly sees, that a fixed order NLO-dijet calculation is not sufficient, 
whereas {\it J2003 set 2} gives a good description of the data.
However, the variable $S$ is  sensitive to  the details of the unintegrated
gluon distribution, as can be seen from the comparison with 
{\it J2003 set 1} and {\it set 3}.
\par
A measurement, aiming to observe deviations from the
collinear DGLAP approach, is the production of jets in the forward (proton)
region. The phase space is restricted to a region of $Q^2 > 5$ \gevsq~and 
$E_T^{jet} > 3.5$ \gev~in the forward region of 
$1.7 <\eta_{jet} < 2.8$ 
with the additional requirement of 
$0.5 < E^2_{{T,jet}}/Q^2 < 2$, a region where the 
 contribution from the evolution in $Q^2$ is small. 
The cross section for forward jet production has been measured by 
H1~\cite{H1fwdjet2} as a function of \xbj , shown in
Fig.~\ref{fjet:cas-nlo} together with predictions from \cascade . 
Also the NLO-dijet~\cite{DISENT}
prediction 
in the collinear approach is shown.
The fixed NLO-dijet calculation falls below the measurement, whereas the
\kt-factorization approach supplemented with CCFM evolution gives a reasonable
description of the data. 

\subsection{Heavy Quark Production at the Tevatron}
\begin{figure}[ht]
\vskip -0.5cm
\centerline{\resizebox{1.1\textwidth}{!}{\includegraphics{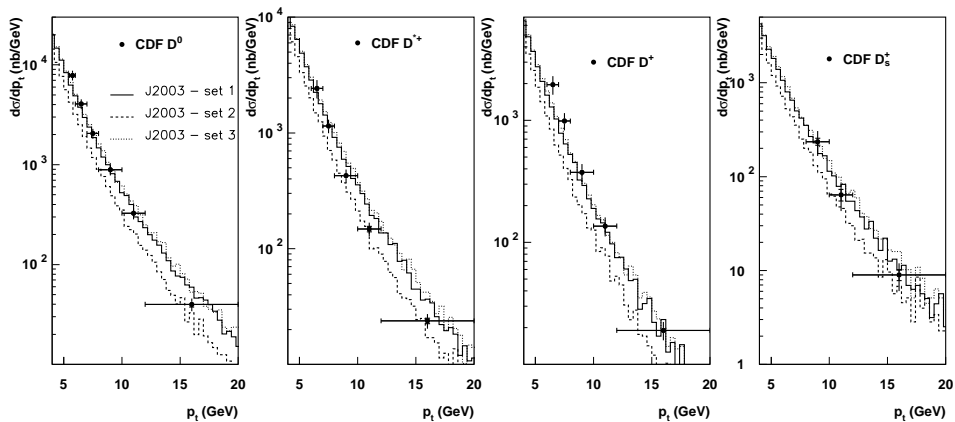}}} 
\vskip -0.5cm
\caption{ Differential cross section for $D$ meson production as measured by 
CDF~\protect\cite{CDFcharm} as a function of the transverse momentum compared to
predictions from \cascade .
\label{charmcdf}} 
\end{figure}

The differential cross section as a function of the transverse momentum of
$D$-mesons  
has been measured in $p\bar{p}$ collisions 
at $\sqrt{s}=1.96$~TeV by the
CDF collaboration~\cite{CDFcharm}. They find the measured 
cross section to be larger than the NLO 
predictions in the collinear factorization approach by about 100 \% at low \pt\ 
and 50 \% at high \pt. 
In Fig.~\ref{charmcdf} the measurement is shown together with
the predictions obtained in \kt-factorization using 
\cascade\ with the CCFM unintegrated gluon density described above. 
For {\it J2003 set 1} and {\it set 3} good agreement for all 
measured charmed mesons is observed.  
\par
The cross section for $b \bar{b}$ production 
in $p\bar{p}$ collision at
$\sqrt{s}=1800$~GeV has been compared
with the prediction of \cascade\ based on the CCFM gluon densities in
\cite{jung-mpla2003}. It is interesting to note that {\it J2003 set 2} which
describes best the data of Fig.~\ref{azim:casc-nlo}, 
predicts a lower cross section for heavy quark production.
\section{Conclusion}
It has been shown, that \kt-factorization and the CCFM evolution of the gluon
density is a powerful tool for the description of hadronic final state
measurements. Improvements of the CCFM splitting function have been discussed.
\par
Jet measurements at HERA, but also measurements of charm and bottom production
at the Tevatron can be reasonably well described,  
whereas calculations performed
in the collinear approach even in NLO have difficulties to describe the data. 
This shows the
advantage of applying \kt-factorization to estimate higher order contributions
to the cross section  but also the
importance of a detailed understanding of the parton evolution process.
\section*{Acknowledgments}
It is a pleasure to thank A.~Bialas, K.~Fialkowski, K.~Golec-Biernat, 
B.~Brzezicka and the whole crew for this interesting and stimulating workshop. 
\raggedright


\begin{thebibliography}{10}

\bibitem{H1F2}
C.~Adloff et~al.,  (2003)\nolinebreak [2]\,.

\bibitem{ZEUSF2}
\mbox{ZEUS} Collaboration;S. Chekanov~et al., {\em Phys. Rev.} {\bf
  D67}\nolinebreak [2]\,(2003)\nolinebreak [2]\,012007.

\bibitem{H1alpha}
C.~Adloff et~al., {\em Eur. Phys. J.} {\bf C19}\nolinebreak
  [2]\,(2001)\nolinebreak [2]\,289.

\bibitem{ZEUSalpha}
J.~Breitweg et~al., {\em Phys. Lett.} {\bf B507}\nolinebreak
  [2]\,(2001)\nolinebreak [2]\,70.

\bibitem{D0comp}
B.~Abbott et~al., {\em Phys. Rev. Lett.} {\bf 82}\nolinebreak
  [2]\,(1999)\nolinebreak [2]\,2457.

\bibitem{DGLAPa}
V.~Gribov, L.~Lipatov, {\em Sov. J. Nucl. Phys.} {\bf 15}\nolinebreak
  [2]\,(1972)\nolinebreak [2]\,438 and 675.

\bibitem{DGLAPb}
L.~Lipatov, {\em Sov. J. Nucl. Phys.} {\bf 20}\nolinebreak
  [2]\,(1975)\nolinebreak [2]\,94.

\bibitem{DGLAPc}
G.~Altarelli, G.~Parisi, {\em Nucl. Phys. {\bf B}} {\bf 126}\nolinebreak
  [2]\,(1977)\nolinebreak [2]\,298.

\bibitem{DGLAPd}
Y.~Dokshitser, {\em Sov. Phys. JETP} {\bf 46}\nolinebreak
  [2]\,(1977)\nolinebreak [2]\,641.

\bibitem{BFKLa}
E.~Kuraev, L.~Lipatov, V.~Fadin, {\em Sov. Phys. JETP} {\bf 44}\nolinebreak
  [2]\,(1976)\nolinebreak [2]\,443.

\bibitem{BFKLb}
E.~Kuraev, L.~Lipatov, V.~Fadin, {\em Sov. Phys. JETP} {\bf 45}\nolinebreak
  [2]\,(1977)\nolinebreak [2]\,199.

\bibitem{BFKLc}
Y.~Balitskii, L.~Lipatov, {\em Sov. J. Nucl. Phys.} {\bf 28}\nolinebreak
  [2]\,(1978)\nolinebreak [2]\,822.

\bibitem{CCH}
S.~Catani, M.~Ciafaloni, F.~Hautmann, {\em Nucl. Phys. {\bf B}} {\bf
  366}\nolinebreak [2]\,(1991)\nolinebreak [2]\,135.

\bibitem{CE}
J.~Collins, R.~Ellis, {\em Nucl. Phys. {\bf B}} {\bf 360}\nolinebreak
  [2]\,(1991)\nolinebreak [2]\,3.

\bibitem{GLR}
L.~Gribov, E.~Levin, M.~Ryskin, {\em Phys. Rep.} {\bf 100}\nolinebreak
  [2]\,(1983)\nolinebreak [2]\,1.

\bibitem{LRSS2}
E.~M. Levin, M.~G. Ryskin, Y.~M. Shabelski, A.~G. Shuvaev, {\em Sov. J. Nucl.
  Phys.} {\bf 53}\nolinebreak [2]\,(1991)\nolinebreak [2]\,657.

\bibitem{CCFMa}
M.~Ciafaloni, {\em Nucl. Phys. {\bf B}} {\bf 296}\nolinebreak
  [2]\,(1988)\nolinebreak [2]\,49.

\bibitem{CCFMb}
S.~Catani, F.~Fiorani, G.~Marchesini, {\em Phys. Lett. {\bf B}} {\bf
  234}\nolinebreak [2]\,(1990)\nolinebreak [2]\,339.

\bibitem{CCFMc}
S.~Catani, F.~Fiorani, G.~Marchesini, {\em Nucl. Phys. {\bf B}} {\bf
  336}\nolinebreak [2]\,(1990)\nolinebreak [2]\,18.

\bibitem{CCFMd}
G.~Marchesini, {\em Nucl. Phys. {\bf B}} {\bf 445}\nolinebreak
  [2]\,(1995)\nolinebreak [2]\,49.

\bibitem{LDCa}
B.~Andersson, G.~Gustafson, J.~Samuelsson, {\em Nucl. Phys. {\bf B}} {\bf
  467}\nolinebreak [2]\,(1996)\nolinebreak [2]\,443.

\bibitem{LDCb}
B.~Andersson, G.~Gustafson, H.~Kharraziha, J.~Samuelsson, {\em Z. Phys. {\bf
  C}} {\bf 71}\nolinebreak [2]\,(1996)\nolinebreak [2]\,613.

\bibitem{smallx_2001}
\mbox{B.~Andersson {\it et al.} [Small x Collaboration]}, {\em Eur. Phys. J.
  {\bf C}} {\bf 25}\nolinebreak [2]\,(2002)\nolinebreak [2]\,77,
  hep-ph/0204115.

\bibitem{Forshaw:1998uq}
J.~R. Forshaw, A.~Sabio~Vera, {\em Phys. Lett.} {\bf B440}\nolinebreak
  [2]\,(1998)\nolinebreak [2]\,141.

\bibitem{Webber:1998we}
B.~R. Webber, {\em Phys. Lett.} {\bf B444}\nolinebreak [2]\,(1998)\nolinebreak
  [2]\,81.

\bibitem{Salam:1999ft}
G.~Salam, {\em JHEP} {\bf 03}\nolinebreak [2]\,(1999)\nolinebreak [2]\,009.

\bibitem{jung_salam_2000}
H.~Jung, G.~Salam, {\em Eur. Phys. J. {\bf C}} {\bf 19}\nolinebreak
  [2]\,(2001)\nolinebreak [2]\,351, \mbox{hep-ph/0012143}.

\bibitem{CASCADEMC}
H.~Jung, {\em Comp. Phys. Comm.} {\bf 143}\nolinebreak [2]\,(2002)\nolinebreak
  [2]\,100, \verb+http://www.quark.lu.se/~hannes/cascade/+.

\bibitem{LDCc}
G.~Gustafson, H.~Kharraziha, L.~L\"onnblad,    Proc. of the Workshop on Future
  Physics at HERA, edited by A.~\mbox{De Roeck}, G.~Ingelman, R.~Klanner.

\bibitem{LDCd}
H.~Kharraziha, L.~L\"onnblad, {\em JHEP} {\bf 03}\nolinebreak
  [2]\,(1998)\nolinebreak [2]\,006.

\bibitem{jung-dis03}
M.~Hansson, H.~Jung, The status of CCFM unintegrated gluon densities, \mbox{DIS
  2003}, St. Petersburg, Russia, hep-ph/0309009.

\bibitem{H1_F2_1996}
\mbox{H1 Collaboration, S. Aid et al.}, {\em Nucl. Phys. {\bf B}} {\bf
  470}\nolinebreak [2]\,(1996)\nolinebreak [2]\,3.

\bibitem{H1_F2_2001}
\mbox{H1 Collaboration, C. Adloff et al.}, {\em Eur. Phys. J. {\bf C}} {\bf
  21}\nolinebreak [2]\,(2001)\nolinebreak [2]\,33.

\bibitem{ZEUS_F2_1996}
\mbox{ZEUS} Collaboration; M. Derrick~et al., {\em Z. Phys.} {\bf
  C72}\nolinebreak [2]\,(1996)\nolinebreak [2]\,399.

\bibitem{ZEUS_F2_2001}
\mbox{ZEUS} Collaboration; S. Chekanov~et al., {\em Eur. Phys. J.} {\bf
  C21}\nolinebreak [2]\,(2001)\nolinebreak [2]\,443.

\bibitem{H1dijet}
\mbox{H1} Collaboration; A. Aktas~et al., Inclusive dijet production at low
  Bjorken-x in deep inelastic scattering, 2003, hep-ex/0310019.

\bibitem{Szczurek}
A.~Szczurek, N.~N. Nikolaev, W.~Schafer, J.~Speth, {\em Phys. Lett.} {\bf
  B500}\nolinebreak [2]\,(2001)\nolinebreak [2]\,254.

\bibitem{DISENT}
S.~Catani, M.~H. Seymour, {\em Nucl. Phys.} {\bf B485}\nolinebreak
  [2]\,(1997)\nolinebreak [2]\,291.

\bibitem{H1fwdjet2}
\mbox{H. Jung, for H1 and ZEUS collaborations}, {\em Nuclear Physics B -
  Proceedings Supplements} {\bf 117}\nolinebreak [2]\,(2002)\nolinebreak
  [2]\,352.

\bibitem{CDFcharm}
\mbox{CDF} Collaboration; D. Acosta~et al., Measurement of prompt charm meson
  production cross sections in p anti-p collisions at s**(1/2) = 1.96-TeV,
  2003, hep-ex/0307080.

\bibitem{jung-mpla2003}
H.~Jung, $k_t$ - factorization and CCFM - the solution for describing the
  hadronic final states - everywhere, 2003, hep-ph/0311249, to be published in
  Mod.Phys.Lett. A.

\end{thebibliography}
\end{document}